\documentclass[10pt,letterpaper]{article}
\pdfoutput=1
\usepackage{cite}
\usepackage{textcomp}
\usepackage{amsmath}
\usepackage{nicefrac}
\usepackage{authblk}
\usepackage{graphicx}

\usepackage{dblfloatfix}

\title{Low-Loss Silicon Platform for Broadband Mid-Infrared Photonics}
\date{}

\author[1,2]{Steven A. Miller}
\author[3,4]{Mengjie Yu}
\author[1,2]{Xingchen Ji}
\author[1]{Austin G. Griffith}
\author[2,5]{Jaime Cardenas}
\author[4]{Alexander L. Gaeta}
\author[2,*]{Michal Lipson}

\affil[1]{School of Electrical and Computer Engineering, Cornell University, Ithaca, New York 14853, USA}
\affil[2]{Department of Electrical Engineering, Columbia University, New York, New York 10027, USA}
\affil[3]{School of Applied and Engineering Physics, Cornell University, Ithaca, New York 14853, USA}
\affil[4]{Department of Applied Physics and Applied Mathematics, Columbia University, New York, New York 10027, USA}
\affil[5]{Currently at: The Institute of Optics, University of Rochester, Rochester, NY 14627}
\affil[*]{Corresponding author: ml3745@columbia.edu}

\begin{document}
\maketitle
\begin{abstract}
\noindent Broadband mid-infrared (mid-IR) spectroscopy applications could greatly benefit from today’s well-developed, highly scalable silicon photonics technology; however, this platform lacks broadband transparency due to its reliance on absorptive silicon dioxide cladding. Alternative cladding materials have been studied, but the challenge lies in decreasing losses while avoiding complex fabrication techniques. Here, in contrast to traditional assumptions, we show that silicon photonics can achieve low-loss propagation in the mid-IR from 3 – 6 $\mu$m wavelength, thus providing a highly scalable, well-developed technology in this spectral range. We engineer the waveguide cross section and optical mode interaction with the absorptive cladding oxide to reduce loss at mid-IR wavelengths. We fabricate a microring resonator and measure an intrinsic quality ($\textit{Q}$) factor of 10$^6$ at wavelengths from 3.5 to 3.8 $\mu$m. This is the highest $\textit{Q}$ demonstrated on an integrated mid-IR platform to date. With this high-$\textit{Q}$ silicon microresonator, we also demonstrate a low optical parametric oscillation threshold of 5.2 mW, illustrating the utility of this platform for nonlinear chip-scale applications in the mid-IR.	
\end{abstract}

\section{Introduction}
Silicon photonics has revolutionized optics in the near infrared spectral range but has had limited impact on optics in the mid-infrared (mid-IR) range, primarily because of the platform's intrinsic material absorption. Within the mid-IR wavelength range, many common molecules relevant to real-world sensing applications have strong absorption signatures, forming the so-called "molecular fingerprint" spectral region \cite{tittel_mid-infrared_2003}. Numerous forms of optical spectroscopy have been developed for biological, chemical, environmental, and industrial applications, with an increasing need for compact and inexpensive sensors enabled by integrated photonics \cite{tittel_mid-infrared_2003}. The advancement of quantum cascade lasers has paved the way for growth of compact, chip-scale mid-IR photonic applications \cite{faist_quantum_1994,hugi_mid-infrared_2012,spott_quantum_2016}. Concurrently, silicon photonics has transformed near-IR telecommunications technologies while leveraging the immense semiconductor fabrication infrastructure of the electronics industry \cite{streshinsky_road_2013,lim_review_2014}. Despite their development in tandem, silicon photonics has long been considered unsuitable for mid-IR applications due to its lack of broad transparency beyond the near-IR range. The transparency window of the silicon waveguide core extends to $\sim$8 $\mu$m wavelength \cite{soref_silicon_2006,milosevic_rib_2009}, whereas the silicon dioxide (SiO$_2$) cladding material begins absorbing strongly around 3.5 $\mu$m \cite{kitamura_optical_2007}. Therefore, few demonstrations of silicon-on-insulator (SOI) waveguides have operated past 2.5 $\mu$m \cite{mashanovich_low_2011,reimer_mid-infrared_2012,milosevic_silicon_2012,nedeljkovic_silicon_2013,khan_silicon--nitride_2013,roelkens_silicon-based_2013,zlatanovic_mid-infrared_2010}. While mid-IR platforms based on crystalline fluorides \cite{lecaplain_mid-infrared_2016,savchenkov_generation_2015}, germanium \cite{soref_mid-infrared_2010,chang_low-loss_2012,malik_germanium--silicon_2013,brun_low_2014,carletti_nonlinear_2015,shen_mid-infrared_2015,younis_germanium--soi_2016}, chalcogenide materials \cite{lin_high-q_2016,eggleton_chalcogenide_2011,carlie_integrated_2010,ma_low-loss_2013,zha_inverted-rib_2014} have been demonstrated with low loss, integrated photonics structures on a silicon chip in this spectral range have yet to be demonstrated.

Recently, efforts have been made to develop mid-IR photonics on a silicon platform based on suspended structures and alternative substrates; however, to date these have not been able to yield low loss optical structures. The challenge of incorporating alternative cladding materials for silicon-based mid-IR photonics lies in achieving broadband transparency while ensuring low propagation loss with minimal fabrication complexity. Removing the oxide cladding to form either a pedestal or fully suspended SOI waveguide can greatly reduce or eliminate the absorption loss, extending the platform's transparency window \cite{singh_silicon--sapphire_2015,singh_mid-infrared_2014,lin_air-clad_2013,cheng_mid-infrared_2012,xia_suspended_2013,chiles_high-contrast_2013,soler_penades_suspended_2014,jiang_compact_2014}. However, this approach can introduce fabrication complexity as well as challenges for the device's thermal stability and low scattering losses. To increase robustness, the under-cladding can instead be replaced by a more transparent material, such as sapphire (transparent out to $\sim$4.5 $\mu$m). Silicon-on-sapphire is commercially available and has been extensively studied \cite{spott_silicon_2010,baehr-jones_silicon--sapphire_2010,li_low_2011,wong_characterization_2012,shankar_integrated_2013,singh_midinfrared_2015,kalchmair_cascaded_2015,zou_mid-infrared_2015,huang_silicon--sapphire_2016}; however, it is a hard material that is difficult to etch, leading to scattering losses. Additionally, Chen et al. have recently demonstrated a platform based on silicon on calcium fluoride (CaF$_2$), which extends the transparency window even further \cite{chen_heterogeneously_2014}. However, this platform introduces significant complexity, requiring completely custom fabrication processing techniques.

\section{Design of Low-loss Mid-IR Platform}
Here, we demonstrate the ability to leverage silicon photonics to achieve broadband low propagation losses below 1 dB/cm in the mid-IR (comparable with the losses achievable in the near-IR spectral range) by tailoring the waveguide cross-sectional geometry. The primary source of loss for mid-IR silicon waveguides comes from absorption of the mode that overlaps partially with the oxide cladding, which surrounds the waveguide core on all sides. The degree of mode overlap with the cladding, particularly for high-index contrast waveguides, can be tailored by varying cross-sectional geometry as well as index contrast, wavelength, and polarization. Here, in order to minimize these losses, we engineer the mode to minimize propagation loss. Fig. 1 shows the simulated absorption losses for different waveguide geometries, including a traditional silicon photonic waveguide for this wavelength with cross-sectional dimensions 500 nm $\times$ 1400 nm (as in \cite{griffith_silicon-chip_2015}), using finite element method (FEM) simulations of optical modes in COMSOL Multiphysics software. We account for the cladding oxide absorption, adapted from literature \cite{kitamura_optical_2007}, which we plot in Fig. 1 (blue curve). The losses are extracted from the imaginary component of the simulated mode effective refractive index, $k$, from which we calculate propagation power loss, $\alpha = 4\pi k/\lambda$. One can see that the loss increases significantly at mid-IR wavelengths as the mode grows larger into the cladding, exceeding 10 dB/cm, which is prohibitive for most waveguiding applications. If we remove the top cladding, we see a modest improvement in loss, limited still by the large remaining overlap with the under-cladding, as seen in the inset mode picture. By using an unconventional air-clad waveguide geometry with 2300 nm $\times$ 4000 nm cross-section, we are able to significantly decrease the loss at higher wavelengths by more than two order-of-magnitude loss improvement over the oxide-clad waveguide. As evident in the inset mode pictures, the optical mode for this geometry overlaps minimally with the oxide under-cladding compared to the traditional cross-section. For this geometry, the propagation loss remains below 1 dB/cm up to a wavelength of 6 $\mu$m (corresponding to a $\textit{Q}$ factor $>$ 10$^5$) and below 0.1 dB/cm up to nearly 5 $\mu$m ($\textit{Q}$ $>$ 10$^6$). This broadband silicon photonics transparency window requires no additional fabrication complexity.
\begin{figure}[t]
	\centering
	\includegraphics[width=\linewidth]{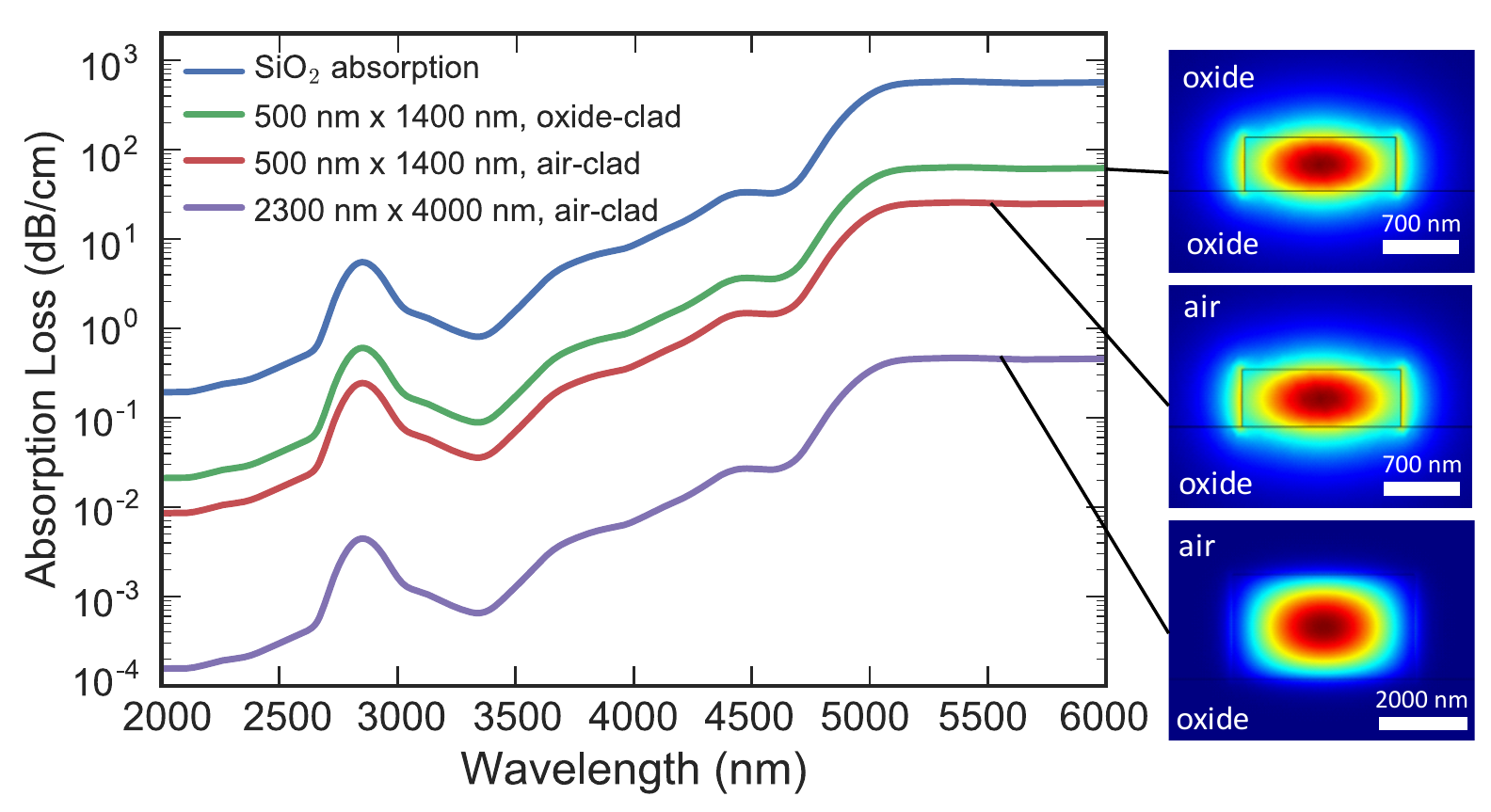}
	\caption{Simulated absorption losses for different waveguide cross-sectional geometries, including a traditional silicon photonic waveguide for this wavelength with cross-sectional dimensions 500 nm $\times$ 1400 nm (as in \cite{griffith_silicon-chip_2015}) as well as our air-clad waveguide with cross-sectional dimensions 2300 nm $\times$ 4000 nm. The cladding oxide absorption, adapted from literature \cite{kitamura_optical_2007}, is plotted in the blue curve. One can see that the air-clad geometry shows a two order-of-magnitude loss improvement compared to the oxide-clad geometry}
	\label{fig1}
\end{figure}

We fabricate an integrated air-clad silicon microring resonator and waveguide, using conventional silicon fabrication processing followed by a chemical treatment of the surfaces to ensure low propagation losses. For our substrate, we use a commercially available SOI wafer from Ultrasil Corporation with a 2.3 $\mu$m top silicon film thickness, and 3 $\mu$m buried oxide thickness. The top silicon film is high resistivity Float-Zone silicon, with a resistivity $>$4,000 $\Omega$.cm, corresponding to an intrinsic doping concentration of $<$ 1$\times$10$^{13}$ cm$^{-3}$. This high resistivity is advantageous for mid-IR wavelengths to minimize losses due to multi-photon absorption. The process involves fabricating unclad photonic structures, integrated with oxide-clad inverse tapers. The oxide-clad inverse tapers guarantee the taper mode to be delocalized for mode-matching to our input laser beam while remaining above the cutoff condition for the propagating waveguide mode.  The taper, which is 200 nm wide at the tip, yields a mode that is designed to overlap with our free-space mid-IR laser beam with a beam width 5 $\mu$m. Note that with an under-cladding thickness of 3 $\mu$m, we can ensure that there is no leakage into the substrate at the input facet by removing the substrate (30-40 $\mu$m) near the edge using xenon difluoride (XeF$_2$), while using photoresist as an etch mask. We pattern the waveguides via electron-beam lithography using ma-N 2410 electron-beam resist, and fully etch the silicon using an inductively-coupled plasma (ICP) dry etching process with a C$_4$F$_8$/SF$_6$ gas mixture. We then deposit 3 $\mu$m of plasma-enhanced chemical vapor deposition (PECVD) silicon dioxide for cladding. In order to remove the cladding above the waveguides and resonators, we pattern a large area above the waveguides and resonators, dry etch the PECVD oxide, leaving 200 nm, and finally remove the remaining cladding oxide in dilute hydrofluoric acid (6:1 H$_2$O:HF) wet etch. In order to clean and smooth the surface, we perform a chemical oxidation surface treatment with piranha acid (3:1 H$_2$SO$_4$:H$_2$O$_2$) and dilute HF, as outlined in \cite{borselli_measuring_2006}. In order to ensure low coupling losses, we then pattern and etch trenches to define the edge facets of the chip \cite{cardenas_high_2014}, and finally dice the wafer into individual chips. Microscope and SEM images of the fabricated waveguide and microring resonator are shown in Fig. 2. 
\setcounter{figure}{1}
\begin{figure}[htbp]
	\centering
	\includegraphics[width=\linewidth]{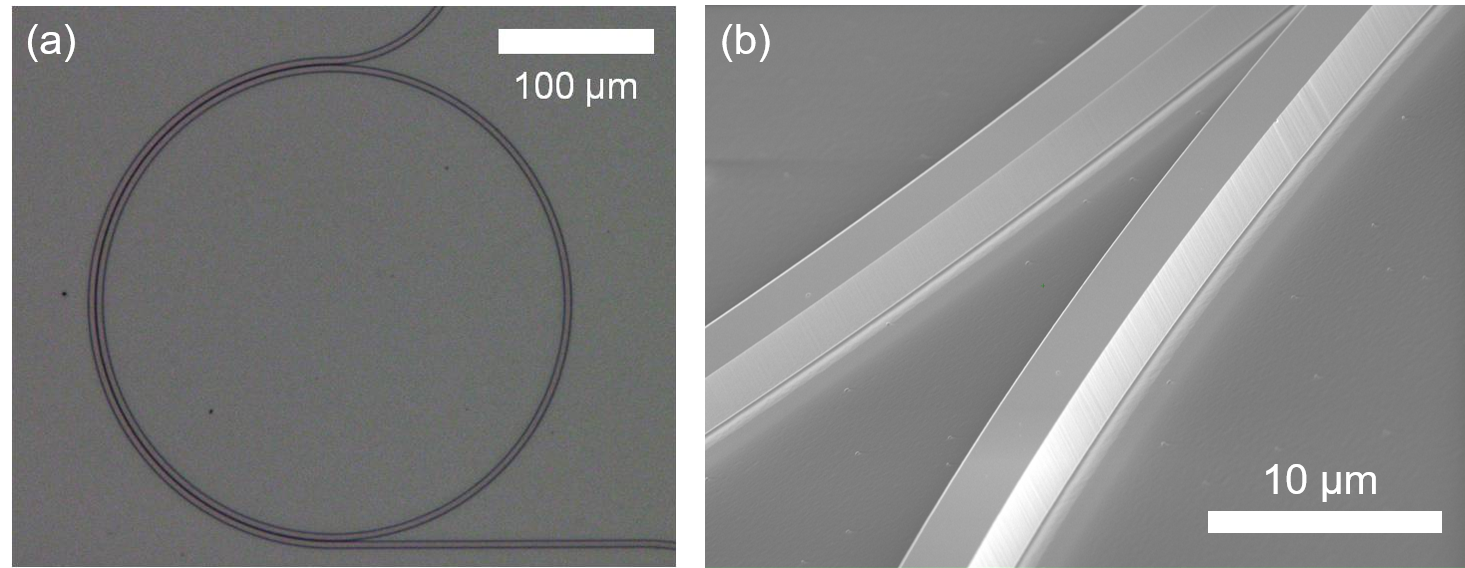}
	\caption{(a) Optical microscope image of air-clad SOI microring resonator. (b) SEM image of air-clad waveguides near the resonator coupling region.}
	\label{fig2}
\end{figure}

\section{High-Q Microring Resonator}
We measure an intrinsic quality factor of 10$^6$ between a wavelength of 3.5 $\mu$m and 3.8 $\mu$m, which is the highest \textit{Q} demonstrated to date in an integrated mid-IR platform. In order to measure the fabricated optical structures, we use as an input mid-IR laser source (Argos Model 2400 CW optical parametric oscillator) that is tunable from 3.2 to 3.8 $\mu$m and has a 100 kHz linewidth. After sending the beam through a variable attenuator and a polarization controller, we couple light into the chip through an aspheric lens. We measure an input coupling loss of 8 dB for TE polarization, likely due to an offset between the designed and fabricated dimensions. We collect light from the chip using another aspheric lens, send it through a polarization filter, and detect using PbSe detector connected to a lock-in amplifier and an optical chopper. To measure the \textit{Q} factor, we modulate the laser wavelength using a triangular wave applied to an internal piezo actuator. We measure a resonance at 3790 nm wavelength (2639 cm$^{-1}$) and we fit to a lorentzian function. To find the intrinsic \textit{Q}, we use the relation:

\begin{equation}\label{eq1}
Q_{\rm i}=\frac{2Q_{\rm l}}{1 \pm \sqrt{T_0}},
\end{equation}

\noindent in which $Q_{\rm i}$ is intrinsic quality factor, $Q_{\rm l}$ is loaded $\textit{Q}$, and $T_0$ is the normalized transmitted power at the resonance wavelength. Assuming an under-coupled cavity condition, we find an intrinsic quality factor of $(1.1 \pm 0.08) \times 10^6$, shown in Fig. 3(a). The laser calibration is detailed below (see Appendix). The output polarization of this resonance is rotated away from TE, indicating some polarization rotation occurring inside the chip, likely from the transition into and out of the air-clad region. We measure intrinsic $\textit{Q}$ at several wavelengths from 3.5 $\mu$m to 3.8 $\mu$m using devices with varying coupling gaps. Measured results are shown in Fig. 3(b) as plotted circles. For comparison, we also measure the intrinsic \textit{Q} of the oxide-clad resonators used in \cite{griffith_silicon-chip_2015}, with a cross-section of 500 nm x 1400 nm. 

\setcounter{figure}{2}
\begin{figure}[htbp]
	\centering
	\includegraphics[width=\linewidth]{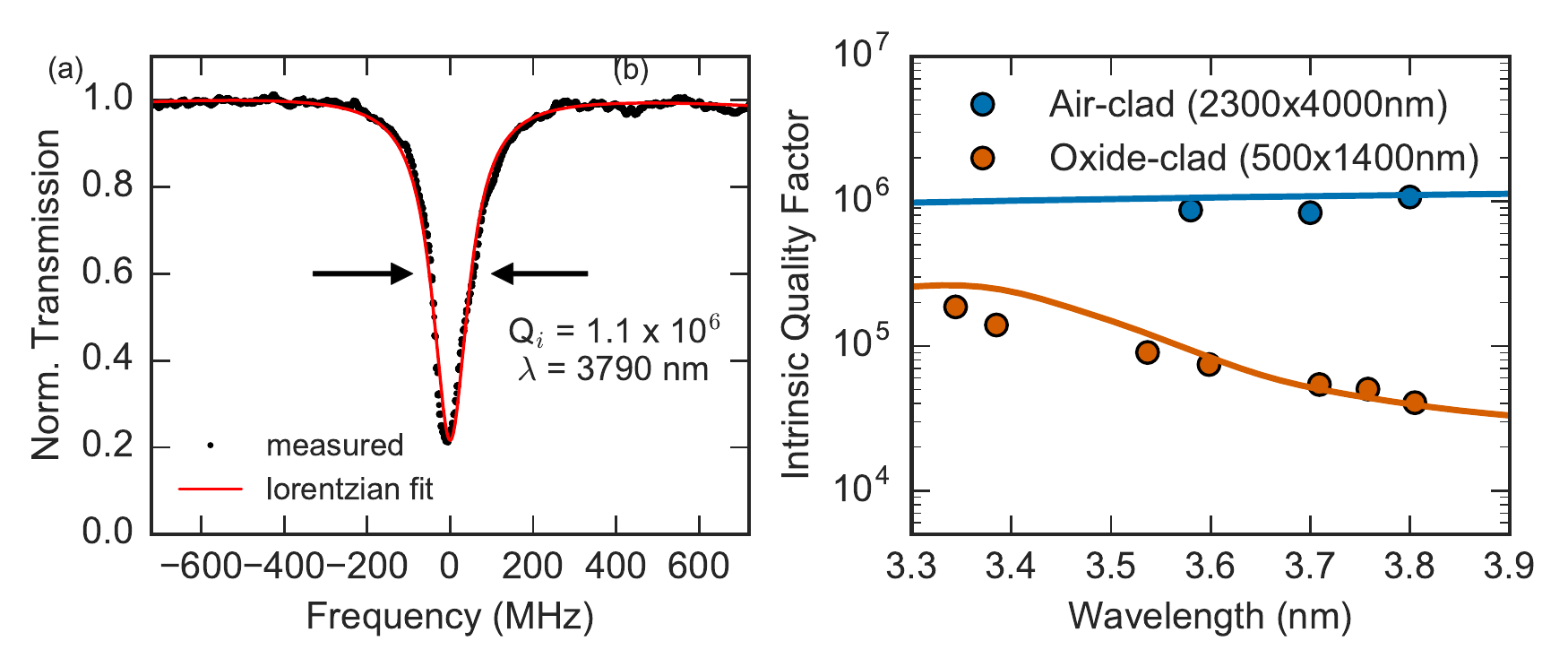}
	\caption{(a) Transmission spectrum of the air-clad microring resonator at 3.8 $\mu$m wavelength. (b) Measured (circles) and simulated (solid lines) intrinsic $\textit{Q}$ as a function of wavelength for our air-clad waveguide geometry (2300 nm $\times$ 4000 nm), and for standard oxide-clad geometry (500 nm $\times$ 1400 nm). There is good agreement between the trends for simulated and measured $\textit{Q}$. One can see that the intrinsic $\textit{Q}$ depends weakly on wavelength between 3 and 4 $\mu$m, in contrast to the standard oxide-clad waveguide geometries.}
	\label{fig3}
\end{figure}


\section{Transparency Window}
For the geometry used here, we show that both the theoretical and measured intrinsic quality factor depends weakly on wavelength between 3.5 and 3.8 $\mu$m, in contrast to the standard oxide-clad waveguide geometries. The theoretical wavelength-dependence of intrinsic quality factor is plotted along with the measured data in Fig 3(b). We evaluate the intrinsic quality factor vs. wavelength by extracting the wavelength-dependent absorption from the FEM simulations and the wavelength-dependent scattering using the well-known theoretical model for waveguide scattering loss introduced by Payne and Lacey \cite{payne_theoretical_1994}. We assume loss due to absorption in the cladding and loss due to sidewall scattering to be the major contributions to the total loss. Therefore, given a representative measured total loss at 3.8 $\mu$m, the scattering loss at 3.8 $\mu$m can be estimated to be 0.2 dB/cm using the relation between $\textit{Q}$ and loss rate:

\begin{equation}\label{eq2}
Q=\frac{2\pi n_{\rm g}}{\lambda\alpha},
\end{equation}

\noindent in which $n_{\rm g}$ is group index \cite{rabiei_polymer_2002}. Using the Payne-Lacey model with our geometry, we can extract the approximate wavelength dependence of the scattering loss to be $\lambda^{-2}$. Thus, quality factor due to scattering loss scales proportional to the wavelength, $\lambda$. In Fig 3(b), plotted with the measured points are the theoretical wavelength dependent quality factor for both measured devices, considering both absorption and scattering. We see a good agreement between the trends for simulated and measured quality factor. The oxide-clad geometry is clearly absorption-limited, whereas our air-clad geometry remains flat across the spectrum, following closely to the theoretically scattering-limited regime.  It is evident that our air-clad geometry achieves a high $\textit{Q}$ independent of the oxide absorption up to an order of magnitude higher than the traditional oxide-clad geometry, which exhibits a strong decrease in $\textit{Q}$ with increasing wavelength.

We show that the waveguide geometry used here extends the transparency window to a wavelength of 6 $\mu$m. In Fig. 4, we plot the wavelength-dependent $\textit{Q}$ due to absorption, scattering, and total loss for our air-clad geometry from 2 $\mu$m to 6 $\mu$m. There is a clear transition from the scattering-limited regime to the absorption-limited regime near 5 $\mu$m wavelength, while for the traditional oxide-clad geometry this transition occurs near 3 $\mu$m wavelength, illustrating a substantial extension of this silicon photonics platform. One can see that throughout the entire spectral region out to 6 $\mu$m, the total intrinsic $\textit{Q}$ remains above 10$^5$ (loss $\le$ 1 dB/cm), and above 10$^6$ (loss $\le$ 0.1 dB/cm) out to 5 $\mu$m, which is well suited for low-loss nonlinear applications. Furthermore, the scattering-limited $\textit{Q}$ is not fundamental, and can be increased through further optimization of side-wall roughness.

\setcounter{figure}{3}
\begin{figure}[htbp]
	\centering
	\includegraphics[width=\linewidth]{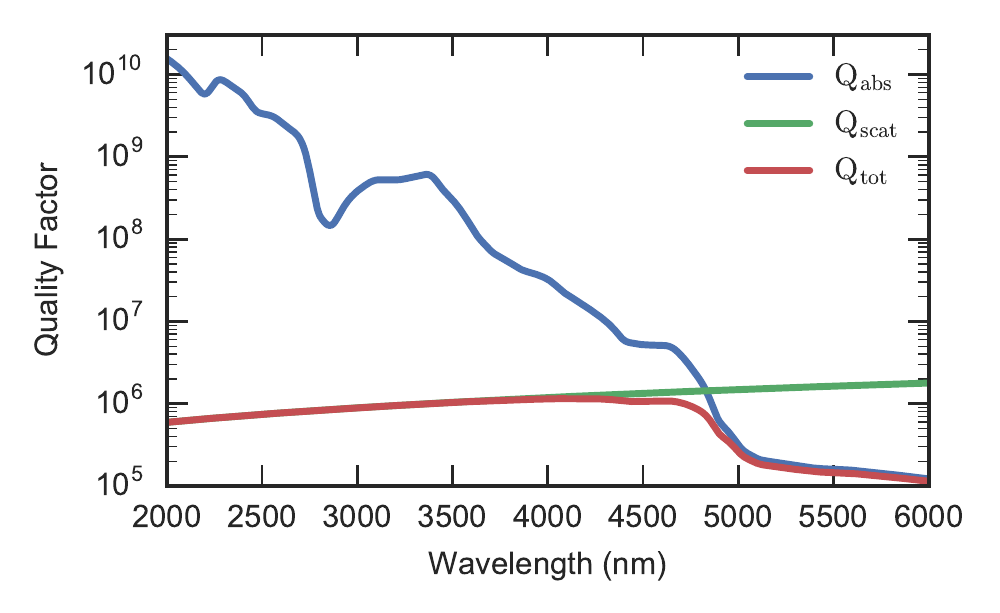}
	\caption{Wavelength dependence of microresonator $\textit{Q}$ (red), due to absorption loss (blue), scattering loss (green). The intrinsic $\textit{Q}$ remains above 10$^{5}$ throughout the whole spectral range between 2 and 6 $\mu$m. Around 5 $\mu$m wavelength, the total intrinsic $\textit{Q}$ transitions from being scattering-limited to absorption-limited.}
	\label{fig4}
\end{figure}

\section{Parametric Oscillation Threshold}
The waveguide geometry used here is highly confined, but maintains the ability to perform dispersion engineering necessary for nonlinear optics applications such as frequency comb generation. Parametric frequency comb generation has been extensively studied in a plethora of integrated platforms, both in the normal dispersion as well as the anomalous dispersion regime \cite{griffith_silicon-chip_2015,delhaye_optical_2007,grudinin_generation_2009,levy_cmos-compatible_2010,liang_generation_2011,wang_mid-infrared_2013,jung_optical_2013,hausmann_diamond_2014,xue_mode-locked_2015,pu_efficient_2016}.  Broadband frequency combs, which are advantageous for mid-IR sensing applications, require anomalous dispersion around the pump wavelength for proper phase-matching \cite{delhaye_octave_2011,okawachi_octave-spanning_2011,kuyken_octave-spanning_2015}. This is usually achieved via waveguide dispersion engineering, which requires certain waveguide confinement and dimensions \cite{turner_tailored_2006}. If the waveguide is too small or too large, the mode resides mostly in the cladding or the core, respectively, and exhibits mostly material dispersion properties that cannot be effectively engineered. Here we show that despite our modified geometry to minimize the waveguide loss, we retain the anomalous dispersion critical for comb generation. Figure 5 shows the dispersion for our air-clad waveguide geometry, which exhibits anomalous dispersion near a pump wavelength of 4 $\mu$m (2500 cm$^{-1}$ wavenumber). This is equidistant in frequency between 3 $\mu$m (3333 cm$^{-1}$) and 6 $\mu$m (1667 cm$^{-1}$), which corresponds to an optimal pump wavelength for broadband comb generation spanning this wavelength range \cite{herr_temporal_2014}. 

\begin{figure}[htbp]
	\centering
	\includegraphics[width=\linewidth]{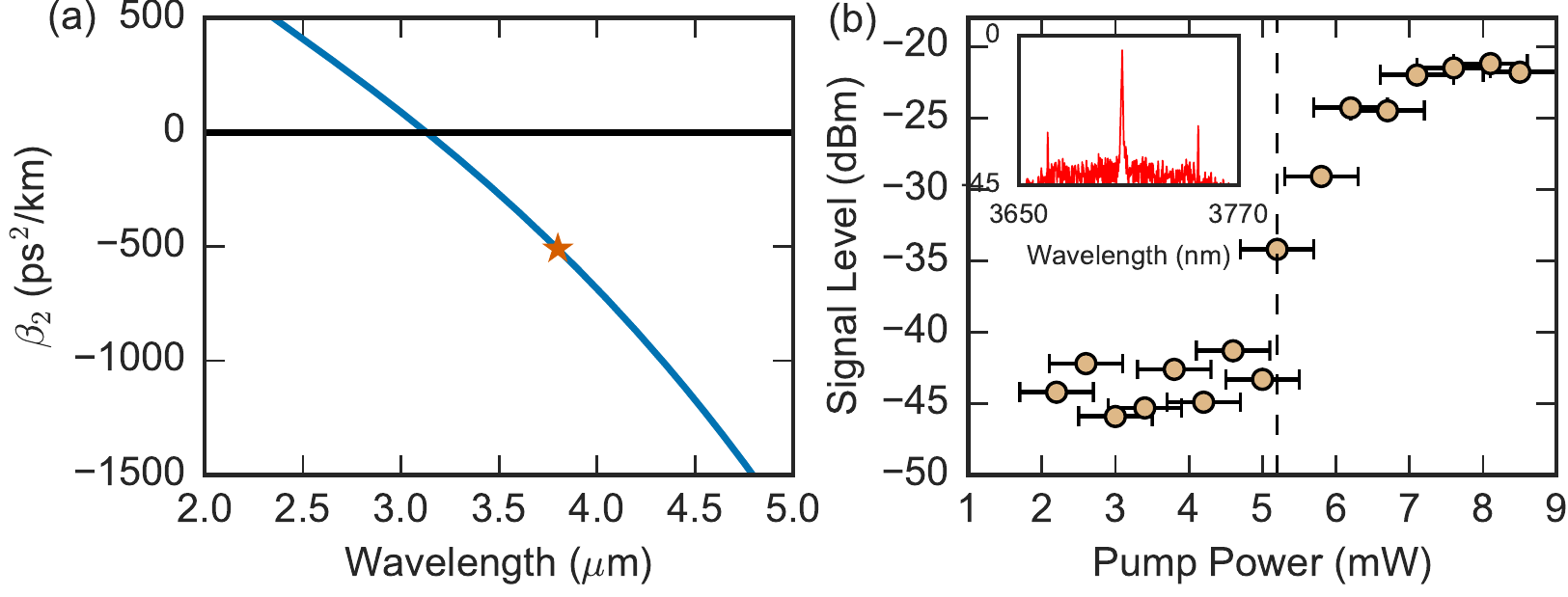}
	\caption{(a) Group velocity dispersion for the waveguide geometry used here (air-clad 2300 nm x 4000 nm). (b) Optical parametric oscillation (OPO) threshold measurement, indicating a 5.2 mW oscillation threshold. Inset shows the OPO sidebands generated just above threshold.}
	\label{fig5}
\end{figure}

We demonstrate a low optical parametric oscillation threshold of 5.2 $\pm$ 0.5 mW using the proposed geometry, which is well within the operating power range of currently available QCLs, enabling future chip-scale applications \cite{spott_quantum_2016}. The parametric oscillation threshold is quadratically dependent on the loaded $\textit{Q}$ factor, making it a critical parameter for frequency comb generation \cite{matsko_optical_2005,gholami_third-order_2011,li_low-pump-power_2012}. In order to measure the threshold power of our microresonator, we use the same experimental setup outlined above, and use a Fourier-transform infrared spectrometer (FTIR) to monitor the mid-IR spectrum. We tune the laser wavelength into a high-$\textit{Q}$ resonance at 3.7 $\mu$m and observe parametric oscillation sidebands generated 10 free-spectral ranges (FSRs) away from the pump wavelength [Fig. 5 (b), inset]. In order to estimate the optical power in the bus waveguide, we assume that the measured output power level is approximately equal to the bus waveguide power inside the chip by collecting the output beam with a lens. We measure an oscillation threshold of 5.2 $\pm$ 0.5 mW [Fig. 5(b)]. 

\section{Conclusion}
In summary, we demonstrate that standard silicon photonics can be instrumental in the development of on-chip photonics in the mid-IR wavelength region. We show a cross-sectional geometry that minimizes the interaction with the absorptive cladding and extends the transparency range out to 6 $\mu$m. This is achieved without addition of complexity in fabrication process. Our demonstrations of a microresonator with an intrinsic $\textit{Q}$ $\sim$ 10$^6$ at 3.8 $\mu$m and an OPO threshold of 5.2 mW illustrate the immediate usefulness of this mid-IR platform for mid-IR frequency comb generation for applications such as on-chip optical spectroscopy \cite{yu_silicon-chip-based_2016,villares_dual-comb_2014}.

\section*{Appendix}
\subsection*{Laser Wavelength Calibration}
For the Argos CW OPO, the piezo actuator acts to tune the pump laser wavelength (centered around 1064 nm) while the signal is held constant, thereby tuning the idler beam at mid-IR wavelengths. We measure the pump wavelength using a wavelength meter to an accuracy of 50 MHz and plot as a function of voltage applied to the piezo element. Performing a linear curve fit given this wavelength error, we extract the fitting error from the covariance matrix, yielding a standard deviation of 500 kHz/V on the slope, which is well within the measurement error.

\section*{Funding Information}
The authors gratefully acknowledge support from Defense Advanced Research Projects Agency (DARPA) under Grant $\#$ W31P4Q-16-1-0002.

\section*{Acknowledgments}

The authors thank Aseema Mohanty, Avik Dutt, and Christopher T. Phare for helpful discussions. This work was performed in part at the CUNY Advanced Science Research Center NanoFabrication Facility.







\end{document}